\documentclass[ba]{imsart}

\RequirePackage{amsthm,amsmath,amsfonts,amssymb}
\RequirePackage[authoryear]{natbib}

\RequirePackage[colorlinks,citecolor=blue,urlcolor=blue,backref=page,backref=page]{hyperref}
\RequirePackage{graphicx}

\pubyear{2025}
\arxiv{2010.00000}
\volume{TBA}
\issue{TBA}
\firstpage{1}
\lastpage{1}

\startlocaldefs
\theoremstyle{plain}

\theoremstyle{definition}

\theoremstyle{remark}

\mathchardef\given="626A

\endlocaldefs

\usepackage{subcaption}
\begin{document}

\begin{frontmatter}
\title{Contributed discussion on: "Model uncertainty and missing data: An Objective Bayesian Perspective"}
\runtitle{Misspecification in Missing data Variable selection}

\begin{aug}
\author[A]{\fnms{Stefan}~\snm{Franssen}\ead[label=e1]{franssen@ceremade.dauphine.fr}\orcid{0000-0001-8870-2135}}
\address[A]{LPSM,
Sorbonne \printead[presep={,\ }]{e1}}

\runauthor{S. Franssen}

\end{aug}

\begin{abstract}
We discuss the effects of model misspecification on variable selection with missing data. 
\end{abstract}



\end{frontmatter}


%

Missing data is a prevalent problem, and has a long history of being studied, see for example~\citet{littleStatisticalAnalysisMissing2019,tsiatisSemiparametricTheoryMissing2006}. 
The theory and methodology has focussed mostly on the frequentist side, and novel Bayesian methods are welcome contributions. 
In this discussion we would like to focus on the imputation distribution.

\noindent
The paper assumed that the model is $\Gamma$-closed.
We argue that with missing data, we should refine our considerations. 
Since we are interested in properties of the model of $Y$ given $X$, we should aim to be robust against misspecification in the model for the distribution of $X$ and consider the distribution of $X$ as a nuisance parameter.

\noindent
The proposed working model is often robust against misspecification in the nuisance parameters.
\citet{lemorvanWhatsGoodImputation2021} studied the effect of imputation rules on the consistency of predictors.
The first effect of misspecification is a loss of efficiency and reliability of the uncertainty quantification.
\citet{kleijnBernsteinVonMisesTheoremMisspecification2012} show that misspecified Bayesian models can be both under- and overconfident, so the credible sets become unreliable.
In extreme cases of misspecification, we can also force a bias in estimates for the $\beta$ parameter, which can lead to inconsistent variable selection.
We will now describe how one can induce bias via misspecification.

\begin{figure}
    \centering
    \begin{subfigure}{0.3 \linewidth}
        \includegraphics[width= 0.9 \linewidth]{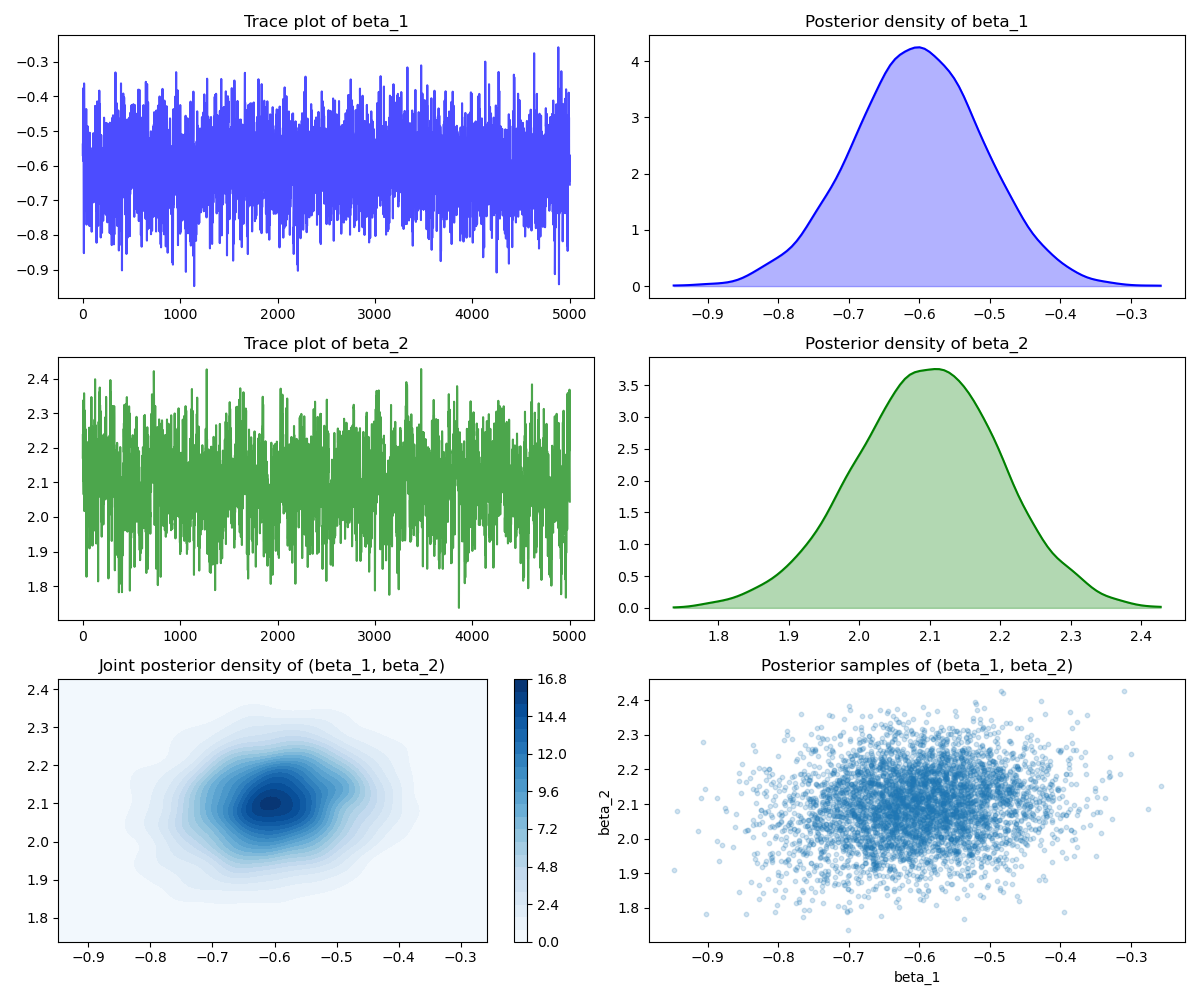}    
        \caption{Simulation 1}%
        \label{fig:sim1}
    \end{subfigure}%
    \begin{subfigure}{0.3 \linewidth}
        \includegraphics[width= 0.9 \linewidth]{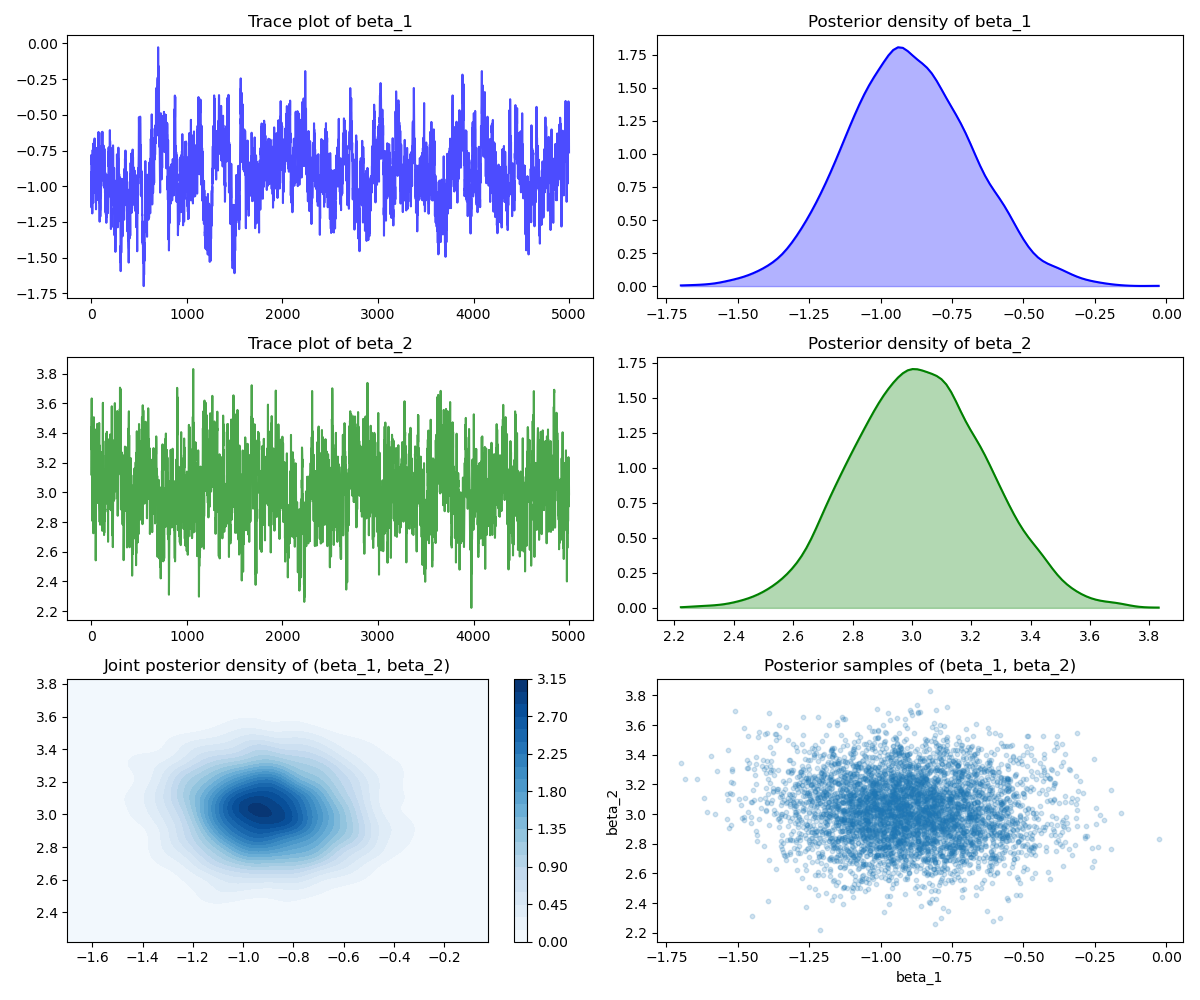}    
        \caption{Simulation 2}%
        \label{fig:sim2}
    \end{subfigure}%
    \begin{subfigure}{0.3 \linewidth}
        \includegraphics[width= 0.9 \linewidth]{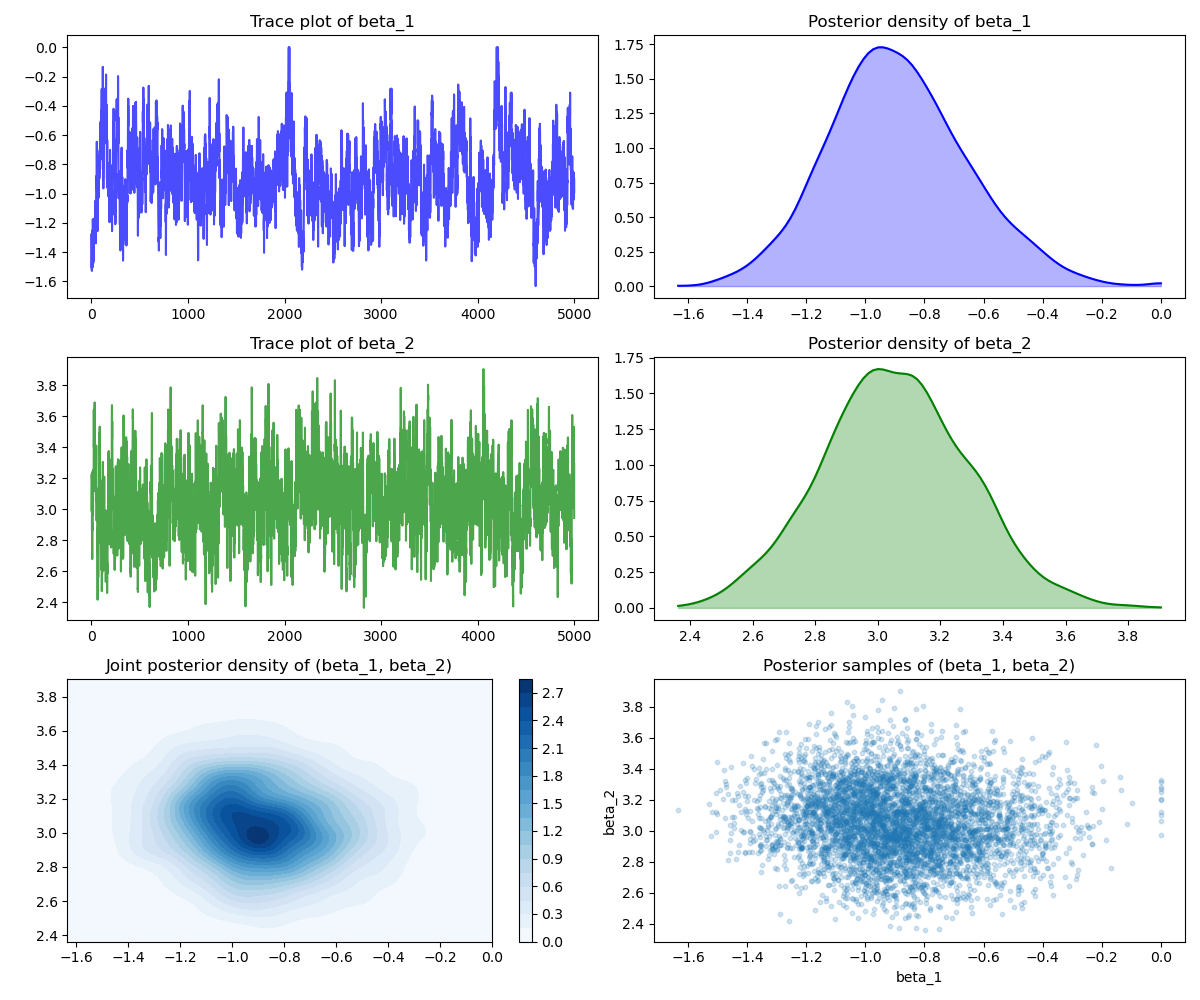}    
        \caption{Simulation 3}%
        \label{fig:sim3}
    \end{subfigure}
\end{figure}

\noindent
The true model will be given by $X_1 \sim N(0,1)$ and $X_2 \given X_1 \sim \exp( e^{-X_1})$.
We will give two examples of simple censoring mechanisms which can lead to misspecification bias:
\begin{itemize}
    \item Censor $X_2$ iff $X_1 < 0$;
    \item Censor $X_2$ iff $|X_1 - Y | < 0.2$.
\end{itemize}
For the first two simulations, we implemented a Zellner $g$-prior without the variable selection procedure using the two examples of censoring mechanisms. 
We also implemented variable selection and used a uniform prior for each of $((), (\beta_1), (\beta_2), (\beta_1, \beta_2))$.
We used $n = 1000$, true variance of $Y$ equal to $1$ and true $\beta_0 = (0, 1)$ in every simulation.
For the posterior density, marginals and trace plots in each simulation, see Figures~\ref{fig:sim1},~\ref{fig:sim2}, and~\ref{fig:sim3}.
In the first simulation, this led to a posterior mean for $\beta$ of $(-0.60, 2.09)$, and the variances of $(\beta_1, \beta_2)$ were $(0.009, 0.01)$.
In the second simulation, this led to a posterior mean for $\beta$ of $(-0.90, 3.03)$, and the variances of $(\beta_1, \beta_2)$ were $(0.05, 0.05)$.
We have included the results for the second selection procedure, the first gave similar results.
The posterior put mass $(0, 0, 0.22, 99.78)$ on the models $(,), (\beta_1,), (\beta_2,), (\beta_1, \beta_2)$.

\noindent
Finally, I would like to pose a list of open questions for the wider Bayesian community.
Can we construct MAR mechanisms that yield bias even when the conditional probability of observing complete data given the observations is bounded from below by a positive constant?
Under what assumptions is the proposed variable selection methodology reliable?
What models allow for an efficient estimation of the true parameters $\beta_0$?
While spike-and-slab (\citet{castilloBayesianLinearRegression2015}) and horseshoe priors (\citet{vanderpasUncertaintyQuantificationHorseshoe2017}) have been studied, frequentist guarantees for Bayesian variable selection with missing data have not yet been explored.
Clarifying these questions would improve the reliability of Bayesian variable selection with missing data.
\bibliographystyle{ba}
\bibliography{MyLibrary}

\end{document}